# Note on Bose-Einstein condensation of photons

Eberhard E. Müller, D - 14621 Schönwalde - Glien
eberhard.mueller@campus.tu-berlin.de

**Abstract**

This paper provides, firstly, a succinct mathematical derivation of Bose-Einstein condensation (BEC) of photons elaborating on previous results in [1,2] including new results on the condensate function and, secondly, applies this framework to consistently explain experimental findings reported in Klaers J., Schmitt, J., Vewinger, F & Weitz, M. [3]. The theoretical approach presented here invites to significantly widen the experimental framework for BEC of photons including three-dimensional photon resonators and thermalization mechanisms different from a dye medium in the cavity.

**Introduction**

Bose-Einstein condensation (BEC) of photons has been demonstrated by Klaers, J., Schmitt, J., Vewinger, F. & Weitz, M. in a specific optical microcavity[3]. Dye molecules filled into the cavity repeatedly absorb and re-emit photons thus providing a thermalization mechanism needed to realize the phase transition. If the number density of the grand-canonical photon gas[4] exceeds a critical value, the excess of photons occupies the ground state of the resonator macroscopically. The design of the cavity makes the photon gas harmonically trapped in an effectively two-dimensional resonator. The photons are formally assigned an effective mass that allows the thermalization mechanism conserving the photon number. Number-conservation is commonly believed a prerequisite[5] to refer to Einstein's original argument for condensation in a monoatomic ideal quantum gas[6].

A low-dimensional harmonic trap system exhibits BEC[7,8]. Applying this finding to the two-dimensional photon gas under consideration presents a conflict. First, the effective mass of the photons depends on the size of the cavity and vanishes for an infinitely large cavity such that the number conservation gets unfounded. Secondly, BEC does not work in a non-relativistic two-dimensional ideal quantum gas[1]. The two-dimensional trap system with BEC qualifies as a weakly interacting quantum gas. Third, a relativistic ideal photon gas admits BEC in three as well as in two dimensions[1]. But this approach is not compatible with photons of non-zero effective mass. It is compatible with the genuine photon concept as a particle with rest mass zero. This approach[1,2], up to now hardly noticed, will be recalled in the following, and applied for BEC of photons in the particular type of an optical microcavity introduced by Klaers, J., Schmitt, J., Vewinger, F. & Weitz, M.[3]. In addition, it will be proved that the photon condensate is accumulated in the center of the resonator. In conclusion the interpretation of the experimental results of Klaers, J. et al.[3] as a BEC of photons in a two-dimensional optical microcavity is supported. The theoretical approach presented here, invites to significantly widen the experimental framework for BEC of photons including three-dimensional photon resonators and thermalization mechanisms

case[6]

different from a dye medium in the cavity.

A prerequisite for a BEC is a grand-canonical ensemble; it may be a constrained one, as in the Einstein case[6], or an unconstrained one where only the mean values of the densities are controlled. To constitute a grand-canonical gas there are at least two independent thermodynamic variables to be identified. For many physical settings the mean particle number density and the mean energy density are employed. However, in the case of a photon gas, the infrared photons give rise to an unrestricted mean photon number density, while the mean energy density remains accessible and controllable thus being recommended as an appropriate thermodynamic variable. As a second thermodynamic variable we shall use the temperature of the photon gas.

Bose-Einstein condensation in an ideal gas of photons, the very first bosons known, implies a serious conceptual problem. In the infinite volume limit of a photon resonator, the energy of the lowest single photon state is zero, a "state without kinetic energy" [6]. Since the rest mass of the photon is zero, a photonic occupation of the lowest energy state seems to have no substance at all. The infinite volume limit is essential to understand Bose-Einstein condensation: A quantum gas in a finite container has discrete energy levels; the infinite volume limit makes the spacing between the levels infinitesimal allowing an energetic redistribution of the gas constituents. This removes an early objection to Einstein's condensation hypothesis raised by Uhlenbeck[9,10].

For a grand-canonical photon resonator with temperature and mean energy density fixed, the chemical potential becomes a function of the resonator volume. In the critical regime the infinite volume limit of the chemical potential tends to zero. A close inspection[1] shows that the infinite volume limit of the grand-canonical mean energy density as a function of the chemical potential differs from the value of the mean energy density with chemical potential being fixed to zero from the very beginning. In mathematical terms: The mean energy density as a function of the chemical potential is not a continuous function. This non-continuity allows a macroscopic occupation of the ground state. A number conserving thermalization is not a necessary precondition for a Bose-Einstein condensation of photons.

**Photon condensation in three and two dimensions**

We consider a photon gas in a finite cavity of volume $V_R$ with reflecting walls. R may denote a characteristic length of the cavity such that $R^3 = V_R$. The lowest eigenvalue $\epsilon_1^R$ of the photon Hamiltonian for the cavity with Dirichlet boundary conditions is strictly positive, $0 < \epsilon_1^R \leq \epsilon_2^R \leq \epsilon_3^R \leq ...$ where $\epsilon_k^R$, k ≥ 2, denote the excited modes. The temperature T and the mean energy density u are assumed to be independent thermodynamic variables of the photon gas. This amounts to assume a deviation of the thermodynamic Planck equilibrium. To establish this, photons are continuously injected into the photon gas where the frequency and the power of the radiation into the cavity is suitably adjusted. Cooling the walls of the cavity the temperature of the photon gas is fixed at a chosen value. As a result the photon flux builds up a steady state of energy with some desired value *u* of the mean energy of the photon gas. The two independent thermodynamic variables constitute a grand-canonical photon gas.

$$\sqrt{-\Delta_R}$$

$\Delta_R$

*u*

The Hamiltonian of free photons in the cavity is given by

(1) $\quad \hbar c \sqrt{-\Delta_R}$

$\Delta_R$ denoting the Dirichlet Laplacian defined in the cavity; $\hbar$ is the reduced Planck constant, c the speed of light. We switch over to an energy spectrum

(2.a) $\quad \lambda_k^R := \epsilon_k^R - \epsilon_1^R$

with zero as the lowest value. Accordingly we introduce a normalized chemical potential

(2.b) $\quad \mu_R \leq 0$.

The integrated spectral density of the photon Hamiltonian is[2]

(3) $\quad F_R(\lambda) := \frac{1}{V_R} \#\{(k, \alpha) \in \mathbb{N} \times \{+1, -1\} : \lambda_{k,\alpha}^R \leq \lambda\}$

$\qquad = \frac{1}{3\pi^2}\left(\frac{\lambda}{\hbar c}\right)^3 - \frac{A_R}{8\pi V_R}\left(\frac{\lambda}{\hbar c}\right)^2 + O\left(\frac{\lambda}{R^2}\right)$

and the spectral density [1,11]

(4) $\quad dF_R(\lambda) = \frac{1}{\pi^2}(\hbar c)^{-3}\lambda^2\, d\lambda - \frac{A_R}{4\pi V_R}(\hbar c)^{-2}\lambda\, d\lambda + O(R^{-2})\, d\lambda;$

$\lambda_{k,\alpha}^R := \lambda_k^R$, $\alpha$ counting the two helicity values of the photons; $A_R$ denotes the surface area of the cavity. The first term gives the bulk contribution to the density, and the second one the surface contribution. In the following we neglect higher orders in $R^{-1}$.

The grand-canonical expectation value of the photon Hamiltonian density gives the mean energy density of the photon gas in the cavity:

(5) $\quad u_R(\beta, \mu_R) = \frac{2}{V_R}\sum_{k=1}^{\infty}\left(\lambda_k^R + \epsilon_1^R\right)\left(e^{\beta(\lambda_k^R - \mu_R)} - 1\right)^{-1}$

The factor 2 sums the helicities, $\beta = 1/(k_B T)$ denotes the inverse temperature of the photon gas; $k_B$ is the Boltzmann constant. Using the spectral density (4), the evaluation of (5) yields[1]:

(6) $\quad u_R(\beta, \mu_R) = \frac{2\epsilon_1^R}{V_R}\left(e^{-\beta\mu_R} - 1\right)^{-1} + \frac{2}{V_R}\sum_{k=2}^{\infty}\left(\lambda_k^R + \epsilon_1^R\right)\left(e^{\beta(\lambda_k^R - \mu_R)} - 1\right)^{-1}$

$\qquad = \frac{2\epsilon_1^R}{V_R}\left(e^{-\beta\mu_R} - 1\right)^{-1} + \frac{2}{V_R}\sum_{k=2}^{\infty}\left(\lambda_k^R + \epsilon_1^R\right)\sum_{n=1}^{\infty}e^{-n\beta(\lambda_k^R - \mu_R)}$

$\qquad = \frac{2\epsilon_1^R}{V_R}\left(e^{-\beta\mu_R} - 1\right)^{-1} + \sum_{n=1}^{\infty}e^{n\beta\mu_R}\frac{2}{V_R}\sum_{k=2}^{\infty}\left(\lambda_k^R + \epsilon_1^R\right)e^{n\beta\lambda_k^R}$

$\qquad \frac{2\epsilon_1^R}{V_R}\left(e^{-\beta\mu_R} - 1\right)^{-1}\quad \sum_{n=1}^{\infty}e^{n\beta\mu_R}\int_{\epsilon_2^R}^{\infty}\left(\lambda + \epsilon_1^R\right)dF_R(\lambda)$

$\qquad u_R(\beta, \mu_R)$

$u_e^R(\beta, \mu_R) \hfill \epsilon_1^R$

$$\frac{2\epsilon_1^R}{V_R}\left(e^{-\beta\mu_R}-1\right)^{-1} \quad \frac{2}{V_R}\sum_{k=2}^{\infty}\left(\lambda_k^R+\epsilon_1^R\right)\sum_{n=1}^{\infty}e^{-n\beta(\lambda_k^R-\mu_R)}$$

$$\sum_{n=1}^{\infty}e^{n\beta\mu_R} \quad \sum_{k=2}^{\infty}\left(\lambda_k^R+\epsilon_1^R\right)e^{n\beta\lambda_k^R}$$

$$=\frac{2\epsilon_1^R}{V_R}\left(e^{-\beta\mu_R}-1\right)^{-1}+\sum_{n=1}^{\infty}e^{n\beta\mu_R}\int_{\epsilon_2^R}^{\infty}\left(\lambda+\epsilon_1^R\right)dF_R(\lambda)$$

The second term of $u_R(\beta,\mu_R)$ represents the excited states of the photon gas; we denote it by $u_e^R(\beta,\mu_R)$. We evaluate the integral and, thereby, neglect the term with $\epsilon_1^R$ which, in the infinite volume limit, tends to 0; the lower integration bound $\epsilon_2^R$ also tends to 0, for $R\to\infty$. The first two terms of the asymptotic expansion of $u_e^R(\beta,\mu_R)$ with respect to R are

(7) $\quad u_e^R(\beta,\mu_R) \sim u_e(\beta,\mu) = \sum_{n=1}^{\infty}e^{n\beta\mu}\left(\frac{1}{n^4}\frac{6}{\pi^2\hbar^3c^3\beta^4}-\frac{1}{n^3}\frac{A_R}{V_R}\frac{2}{4\pi\hbar^2c^2\beta^3}\right)$

$u_e^R(\beta,0)$ is the mean energy density of black body radiation. It turns out to be the critical mean energy density for the photon condensation. The asymptotic expansion gives the bulk contribution $u_{\text{crit}}^{\text{bulk}}(\beta)$ and the surface contribution $u_{\text{crit}}^{\text{surface}}(\beta)$:

(8.a) $\quad u_{\text{crit}}^{\text{bulk}}(\beta) := u_e^{R,\text{bulk}}(\beta,0) = \frac{6}{\pi^2\hbar^3c^3\beta^4}g_4(1)$

(8.b) $\quad u_{\text{crit}}^{\text{surface}}(\beta) := u_e^{R,\text{surface}}(\beta,0) = \frac{2}{4\pi\hbar^2c^2\beta^3}g_3(1)$

where

(9) $\quad g_p(z) := \sum_{n=1}^{\infty}\frac{z^n}{n^p}, \quad \zeta(p) = g_p(1).$

$\zeta$ is the Riemannian zeta function; $g_4(1) = \zeta(4) = \pi^4/90$; $g_3(1) = \zeta(3) = 1.20206...$

Given a temperature $\beta$, and a value $\underline{u}$ of the mean energy density. Then the chemical potential $\mu_R$ is a dependent variable determined by the equation

(10) $\quad u_R(\beta,\mu_R) = \underline{u}$

In the bulk approximation, $\mu_R\to\mu$ where the thermodynamic limit $\mu$ is a unique solution [12] of

(11) $\quad u_e^{\text{bulk}}(\beta,\mu) := \frac{6}{\pi^2\hbar^3c^3\beta^4}g_4(e^{\beta\mu}) = \underline{u} \quad \text{if } \underline{u}\le u_{\text{crit}}^{\text{bulk}}(\beta), \text{ and}$

(12) $\quad \mu=0 \quad \text{if } \underline{u}>u_{\text{crit}}^{\text{bulk}}(\beta).$

(12) represents the condensation regime. In this regime the mean energy density of the condensate is given by

(13) $\quad \underline{u}_1 := \underline{u} - u_{\text{crit}}^{\text{bulk}}(\beta) \text{ if } \underline{u}>u_{\text{crit}}^{\text{bulk}}(\beta). \text{ If } \underline{u}\le u_{\text{crit}}^{\text{bulk}}(\beta), \text{ we set } \underline{u}_1:=0.$

$\mu_R$

$$\frac{2\epsilon_1^R}{V_R}\left(e^{-\beta\mu_R}-1\right)^{-1} \quad \frac{2\epsilon_1^R}{R^3}(1-\beta\mu_R+...-1)^{-1} \quad \underline{u}_1$$

$u_1 \quad \underline{u} - u_{\text{crit}}^{\text{bulk}}(\beta) \quad \underline{u} \quad u_{\text{crit}}^{\text{bulk}}(\beta) \quad \underline{u} \quad u_{\text{crit}}^{\text{bulk}}(\beta) \quad u_1:$

The excess energy (13) is absorbed by the ground state represented by the first term on the right hand side in equation (6). This implies the convergence rate of the chemical potential $\mu_R$ when approaching zero:

$$\frac{2\,\epsilon_1^R}{V_R}\left(e^{-\beta\mu_R} - 1\right)^{-1} = \frac{2\,\epsilon_1^R}{R^3}(1 - \beta\mu_R + \ldots - 1)^{-1} = u_1 \Rightarrow$$

(14) $\qquad \mu_R \sim \dfrac{2\,\beta\,\epsilon_1^R}{R^3} \sim \dfrac{1}{R^4}$

since $\epsilon_1^R \sim \frac{1}{R}$. To put the chemical potential zero before performing the infinite volume limit, or to let $\mu_R$ converge to zero in the infinite volume limit procedure (14), that makes a decisive difference giving rise to the condensation.

To consider a two-dimensional ideal photon gas, we refer to the asymptotic expansion of the mean energy density (7). We drop the bulk term and focus on the two-dimensional surface term. Also, higher orders in $R^{-1}$ are neglected. The independent thermodynamic variables are the mean energy surface density $u^s$ (with the unit $J/m^2$), and the inverse temperature $\beta$. We follow the reasoning as in the bulk case. The chemical potential $\mu$ is a unique solution of

(15) $\qquad u_e^{\text{surface}}(\beta, \mu) := \dfrac{2}{4\pi\hbar^2 c^2 \beta^3} g_3(e^{\beta\mu}) = \underline{u^s} \quad \text{if } \underline{u^s} \leq u_{\text{crit}}^{\text{surface}}(\beta), \text{ and}$

(16) $\qquad \mu = 0 \quad \text{if } \underline{u^s} > u_{\text{crit}}^{\text{surface}}(\beta).$

If the value $\underline{u^s}$ lies in the critical regime (16), the mean energy surface density of the condensate emerges spontanously and is given by

(17) $\qquad u_1^s := \underline{u^s} - u_{\text{crit}}^{\text{surface}}(\beta). \text{ If } \underline{u^s} \leq u_{\text{crit}}^{\text{surface}}(\beta), \text{ we set } u_1^s := 0.$

The excess energy (17) occupies the ground state.

The asymptotic expansion of the critical mean energy density $u_e^R(\beta, 0)$ up to second order can be read off from (7) and is given by

(18) $\qquad u_{\text{crit}}(\beta) = u_{\text{crit}}^{\text{bulk}}(\beta) - \dfrac{A_R}{V_R} u_{\text{crit}}^{\text{surface}}(\beta)$

$$= \dfrac{6}{\pi^2 \hbar^3 c^3 \beta^4} g_4(1) - \dfrac{A_R}{V_R} \dfrac{2}{4\pi\hbar^2 c^2 \beta^3} g_3(1)$$

Accordingly the total critical energy of the finite photon gas up to second order is

(19) $\qquad U_{\text{crit}}(\beta) = V_R \dfrac{6}{\pi^2 \hbar^3 c^3 \beta^4} g_4(1) - A_R \dfrac{2}{4\pi\hbar^2 c^2 \beta^3} g_3(1)$

$R_{\text{curv}} \qquad\qquad D_0 \qquad\qquad\qquad {}^3,$

$V_R \quad \pi D_0^2(R_{\text{curv}} - D_0/3)$

$$U_{crit}^{nb}(\beta) \quad V_R \frac{6}{\pi^2 \hbar^3 c^3 \beta^4} g_4(1) - A_R \frac{2}{4\pi\hbar^2 c^2 \beta^3} g_3(1)$$

## Optical microcavity

Applying this formalism to the case of a two-dimensional optical microcavity, it is possible to calculate the critical power of radiation inside the cavity to induce condensation. Referring to the paper of Klaers, J. et al. [3], two curved mirrors, with radius of curvature $R_{curv} = 1$ m and central distance $D_0 = 1.46$ μm, define the geometry of the optical microcavity. The data imply a volume $V_R = \pi D_0^2 (R_{curv} - D_0/3) = 6.70 \cdot 10^{-12}$ $m^3$, and a surface area $A_R = 2\pi R_{curv} D_0 = 9{,}17 \cdot 10^{-6}$ $m^2$. For room temperature of 300 K, the numerical value of the total critical energy (19) is

$$(20) \quad U_{crit}(300\,K) = V_R \cdot 6.1282 \cdot 10^{-6}\,J/m^3 - A_R \cdot 1.3601 \cdot 10^{-11}\,J/m^2$$

$$= 4.11 \cdot 10^{-17}\,J - 12.47 \cdot 10^{-17}\,J$$

$$= -8.36 \cdot 10^{-17}\,J$$

The surface term dominates the bulk term by a factor of three qualifying this microcavity as an approximately two-dimensional system. The minus-sign of the surface-energy accounts for the Dirichlet boundary. To compare the theoretical values (20) with the experimental results in [3], the critical energy of the photon gas has to be related with the critical power inside the cavity. The power inside the cavity consists of the contribution from the photon gas, and the contribution from the pumped dye molecules forming a thermodynamic reservoir. The latter part is about fifty times larger than the contribution from the photon gas (see [3], same notation):

$$(21) \quad N_{exc}/N_{ph} = \tau_{exc}/\tau_{ph} = 1\text{ ns} / 20\text{ ps} = 50$$

$N_{ph}, \tau_{ph}$ denote the average number of the photons in the resonator and the average time between emission and absorption respectively, $N_{exc}$ and $\tau_{exc}$ the number of molecular excitations and their electronic lifetime in the resonator respectively. A characteristic length of the microcavity is given by the ratio of volume to surface, $l_0 = V_R/A_R \approx D_0$. Focussing on the surface term in (20), $U_{crit\,surf}(300\,K) = 12.47 \cdot 10^{-17}\,J$, we get the critical power of the radiation in the two-dimensional photon gas:

$$(22) \quad P_{crit}(300\,K) = (1+50)\,U_{crit\,surf}(300\,K)/(l_0/c) = 1.31\text{ W}.$$

The theoretical value lies within the tolerance of the experimental value $P_{c,exp} = (1.55 \pm 0{,}60)$ W of Klaers, J. et al. [3].

## The localization of the condensate

Now we determine the explicit form of the condensed state. The states of an ideal grand-canonical photon gas in a resonator are given by the Hilbertspace vectors of the symmetric Fock space over the single photon Hilbertspace $\mathsf{H}$,

$$(23) \quad \mathcal{F}(\mathsf{H}) = \oplus_{n=0}^{N} \mathsf{H}^n$$

$$\mathsf{H}^n$$

$$L_1 \quad L_2 \quad L_3$$

$$\frac{-L_i}{2} \quad x_i \quad \frac{L_i}{2}$$

$$\mathbf{H}^n$$

N denoting the total number of photons, and $\mathbf{H}^n$ the symmetrized n-fold direct product of $\mathbf{H}$. To derive the condensed state of the photon gas, we refer to a 3-dimensional parallelepiped with edges $L_1, L_2, L_3$:

(24) $\quad \dfrac{-L_i}{2} \leq x_i \leq \dfrac{L_i}{2}$, i = 1, 2, 3;

the 2-dimensional case, and the case of cavity geometries different from a parallelepiped follow accordingly. The ground state of the photon gas occupied by $N_1$ photons with energy $\epsilon_1^R$, the lowest energy eigenvalue of the cavity (24) with Dirichlet boundaries supposed, is given by

(25) $\quad \prod_{i=1}^{3} \cos\left(\dfrac{\pi}{L_i} x_{1,i}\right) \ldots \cos\left(\dfrac{\pi}{L_i} x_{N_1,i}\right) \in \mathbf{H}^{N_1}$.

To evaluate (25) in the condensation regime, it is crucial to observe that the condensate does not contribute to the grand-canonical entropy density $s_R$. The entropy density $s_R$ is given by the energy density $u_R$, the photon density $\rho_R$ multiplied by the chemical potential $\mu_R{}^*$, and the radiation pressure $p_R$ [1]:

(26) $\quad s_R(\beta, \mu_R{}^*)$

$= \dfrac{2}{V_R} \sum_{k=1}^{\infty} k_B \left\{ \left(\beta \epsilon_k^R - \beta \mu_R{}^*\right)\left(e^{\beta(\epsilon_k^R - \mu_R{}^*)} - 1\right)^{-1} + \log\left(e^{\beta(\epsilon_k^R - \mu_R{}^*)} - 1\right)^{-1} \right\}$

$= k_B \beta \{u_R(\beta, \mu_R{}^*) - \mu_R{}^* \rho_R(\beta, \mu_R{}^*) + p_R(\beta, \mu_R{}^*).$

In this thermodynamic relation for the entropy density the non-normalized chemical potential $\mu_R{}^* \leq \epsilon_1$ has to be used, and the non-normalized energy spectrum as well. The non-normalized $\mu_R{}^*$ and the normalized $\mu_R$ (2.b) are connected as follows: $\mu_R = \mu_R{}^* - \epsilon_1^R$. Therefore (26) can be written as

(26') $\quad s_R(\beta, \mu_R) = k_B \beta \{u_R(\beta, \mu_R) - (\mu_R + \epsilon_1^R) \rho_R(\beta, \mu_R) + p_R(\beta, \mu_R)\}.$

The term $\epsilon_1^R \rho_R(\beta, \mu_R)$ subtracts the ground state contribution in $u_R(\beta, \mu_R)$. In the thermodynamical limit with $\beta$ and $\underline{u}$ as independent variables, we get the result

(27) $\quad \lim_{R \to \infty} s_R(\beta, \mu_R(\beta, \underline{u})) =: s(\beta, \mu(\beta, \underline{u})) = s_e(\beta, \mu(\beta, \underline{u}))$

where $s_e$ comprises the thermodynamic limit with the excited modes k ≥ 2 in (26). The result (27) includes the condensation regime: For the condensation regime $\underline{u} \geq u_{\text{crit}}^{\text{bulk}}(\beta)$, the chemical potential is zero, and we get

(28) $\quad s(\beta, \mu(\beta, \underline{u})) = s_e(\beta, 0) = \dfrac{4}{3} k_B \beta \, u_{\text{crit}}^{\text{bulk}}(\beta)$, for $\underline{u} \geq u_{\text{crit}}^{\text{bulk}}(\beta)$.

Equation (26') corrects equation (10) in the paper [1]. Consequently equation (27) in [1] has to be replaced by equation (27) above, and equation (40.b) in [1] by equation (28)

---

$$u_{\text{crit}}^{\text{bulk}}(\beta)$$

above [13]. Beyond this correction, details of the above calculations can be taken from reference [1].

If we increase $\underline{u}$ beyond $u_{crit}^{bulk}(\beta)$, the entropy density (28) remains constant while the energy increase builds up the condensate. The condensate does not contribute to the entropy, the entropy of the condensate is zero. This means that, in the condensation regime, the ground state is not a mixture of random phases (25) but a pure state, with identical phases for the cosine-functions. This observation implies the following evaluation of (25):

(29) $\quad \prod_{i=1}^{3}\left(\cos\left(\frac{\pi}{L_i}x_i\right)\right)^{N_1} \in \mathsf{H}^{N_1}.$

Expression (29), denoted by $f_1^{N_1}(x_1, x_2, x_3)$, gives the spatial condensate distribution. In the idealization $N_1 \to \infty$, the spatial distribution $f_1$ of the condensate is

(30) $\quad f_1(x_1, x_2, x_3) = \begin{cases} 1 & \text{for } x_i = 0, \; i = 1, 2, 3 \\ 0 & \text{for } 0 < |x_i| \le \frac{L_i}{2}, \; i = 1, 2, 3 \end{cases}$

At any point $(x_1, x_2, x_3)$ outside the center, $f_1^{N_1}(x_1, x_2, x_3)$ forms a bounded number sequence strictly monotonic decreasing with respect to $N_1$ which implies the result (30). (Compare e.g. [14].) The following graphics visualizes the convergence rate of (29) for the $x_1$ component for $N_1 = 5, 100, 5000$; the last case shows the sharpest distribution.

### Graphics

$x_1$ component of the spatial condensate distribution $f_1^{N_1}$, for $N_1 = 5, 100, 5000$ (starting from left).

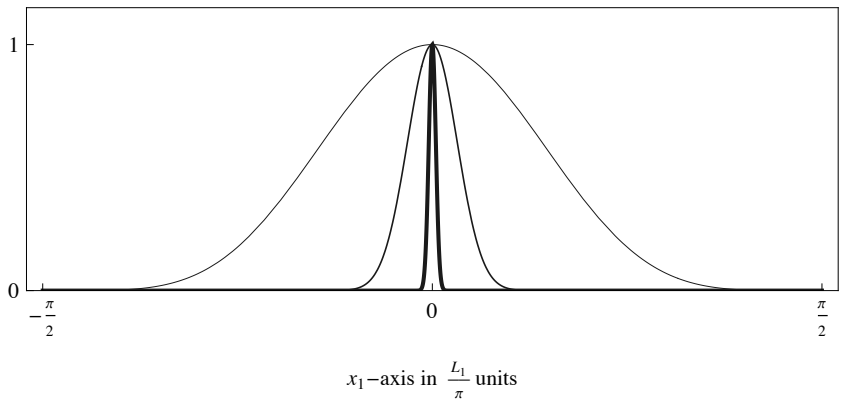

$x_1$-axis in $\frac{L_1}{\pi}$ units

### Discussion

Bose-Einstein condensation in an ideal photon gas can be realized in three and in two dimensions. The analytical framework developed above, allows a formally consistent and quantitatively accurate description of the experimental results of Klaers, J., Schmitt, J., Vewinger, F. & Weitz, M. for a two-dimensional microcavity[3]. In particular, it could be proved that the photon condensate is localized at the center of the cavity, in line with the observation. The proof explains the robustness of the central localization of the condensate against a spatially displacement of the pump beam, as noticed in [3]. The central localization of the photon condensate makes clear that, in the idealization of an infinite number of (infrared) photons with infinitesimally small energy, there is no contribution of the condensate to the radiation pressure, in accordance with the corresponding proof in [1].

Photon condensation transforms photons from higher frequencies to lower frequencies. At the same time the condensate builds up a state of high order. This offers technical applications for photovoltaic energy conversion and energy storage, for new electromagnetic radiation sources, and for photonics.

On the most fundamental level, as described in this paper, the condensate represents stationary energy. Hence, according to Einstein's equivalence of energy and mass, it has to be associated with a non-zero rest mass; the fundamental criterion for matter.

$$^1;$$

The photon condensate does not contribute to the grand-canonical entropy density.